# DESIGN, FABRICATION AND CHARACTERIZATION OF A PIEZOELECTRIC MICROGENERATOR INCLUDING A POWER MANAGEMENT CIRCUIT

*M. Marzencki, Y. Ammar and S. Basrour*

TIMA laboratory, 46, avenue Félix Viallet, 38031 Grenoble FRANCE
E-mail: marcin.marzencki@imag.fr

## ABSTRACT

We report in this paper the design, fabrication and experimental characterization of a piezoelectric MEMS microgenerator. This device scavenges the energy of ambient mechanical vibrations characterized by frequencies in the range of 1 kHz. This component is made with Aluminum Nitride thin film deposited with a CMOS compatible process. Moreover we analyze two possible solutions for the signal rectification: a discrete doubler-rectifier and a full custom power management circuit. The ASIC developed for this application takes advantage of diodes with very low threshold voltage and therefore allows the conversion of extremely low input voltages corresponding to very weak input accelerations. The volume of the proposed generator is inferior to $1mm^3$ and the generated powers are in the range of 1µW. This system is intended to supply power to autonomous wireless sensor nodes.

## 1. INTRODUCTION

Nowadays, a great effort is devoted to the development of Self Powered Micro Systems (SPMS). These devices are commonly used as nodes in Wireless Sensor Networks, where small size and extended autonomy are essential [1]. Up to now, the only solution was to use an electrochemical battery that would supply power to the system. Nevertheless, this solution has a main drawback – the life span of the device is directly linked with the capacity of the energy reservoir and therefore with its size. A trade off has to be made between the miniaturization and the longevity of a device. Ambient power harvesting is a possible breakthrough in this domain. Several research teams have already analysed this subject and it has been found that mechanical vibrations propose very interesting power densities [1]. Furthermore, this kind of energy can be transferred to the device by the means of a simple mechanical coupling. We propose in this paper such an approach consisting in an electromechanical transducer using the piezoelectric effect to convert mechanical vibrations into useful electrical energy. We implement these devices using microfabrication techniques. These MEMS power generators deliver very small powers (in the nW range) at voltages often inferior to 200mV. To overcome the problem of rectification of such weak amplitude AC signals we designed and fabricated diodes with ultra low threshold voltage. These components have been used in a voltage multiplier (VM) which boosts and rectifies the voltage provided by the piezoelectric micro generator. The use of this system was compared with a standard approach using discrete Schottky diodes, similar as presented for a macro prototype by M. Ferrari et al. [2].

The system is used to charge a storage capacitor from which energy is delivered for one cycle of operation of a very low power wireless sensor node. All components of the system are created using CMOS compatible batch microfabrication techniques. In the future it can be realised as a System On a Package (SoP) or even System On a Chip (SoC) in order to reduce its size and cost.

## 2. MICROGENERATOR

### 2.1. Fabrication process

The presented system incorporates a MEMS micro power generator (µPG) that uses the piezoelectric effect for converting the energy of ambient mechanical vibrations into useful electrical energy. The structures are fabricated using an SOI wafer. The Aluminium Nitride thin layer is deposited directly on the heavily doped top silicon layer that serves as the bottom electrode for the piezoelectric capacitor. The piezoelectric layer is sputtered at low temperature and then wet etched in order to define the shape of the mobile structure and the openings for the bottom electrode contact. Then the top aluminium layer is deposited and patterned. Afterwards the top and bottom silicon layers are etched using Deep Reactive Ion Etching (DRIE) in order to create the mobile structure. The use of the AlN layer makes the process compatible with CMOS fabrication because of the lack of high temperature

*Marcin Marzencki, Yasser Ammar and Skandar Basrour*

*Design, fabrication and characterization of a piezoelectric microgenerator including a power management circuit*

treatment. Furthermore, this material is relatively easy to deposit and does not need to be polarised in order to gain piezoelectric properties. Thanks to these advantages, it is a good candidate for industrialisation.

The process employed imposes that the seismic mass thickness is equal to the wafer thickness and the cantilever beam thickness is equal to the top silicon layer thickness. Furthermore the AlN layer thickness was fixed at 1µm. The other dimensions of the structure were optimised using the previously developed models [3]. The dimensions are as follows: the seismic mass measures 800µm by 800µm and the beam is 400µm long and 800µm wide. The length of the beam was imposed by the minimal distance between walls that guarantees uniform back side DRIE etching. The Figure 1 shows a micro photograph of the realised µPG.

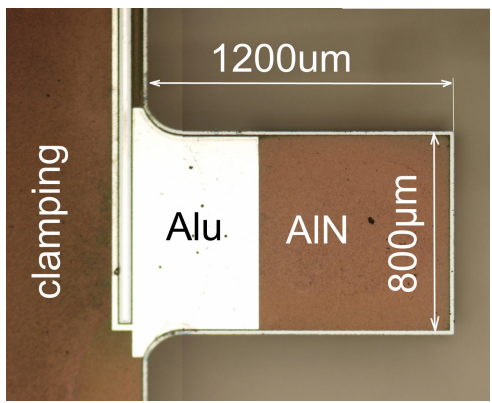

*Figure 1: Micro photograph of the piezoelectric µPG.*

## 2.2. FEM simulations

We have simulated the behavior of this device using ANSYS[TM] software. The FEM model is presented in the Figure 2.

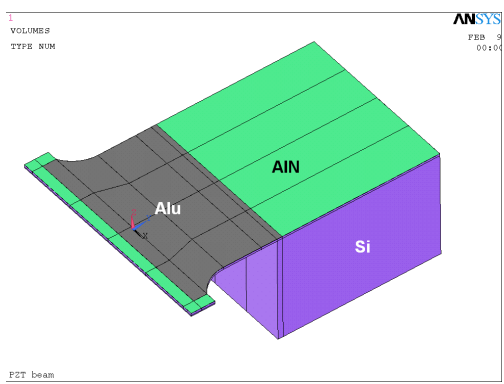

*Figure 2: FEM model of the cantilever multilayer beam with a seismic mass at its end.*

We have made simulations including the parasitic capacitances created between each top metallization and the top silicon layer that serves as the bottom electrode for the piezoelectric capacitor. We have also calculated the quality factor of the system taking into account the quality factor of the AlN layer ($Q_{AlN} = 120$) and the wiscous damping [4]. The silicon layer was considered to be perfect ($Q = 10^5$) in comparison with the piezoelectric layer. The resulting damping coefficient for the structure is equal to 0.145%. The results shown in the Figure 3 give output powers of about 11nW at 0.2g at optimal resistance of 1MΩ.

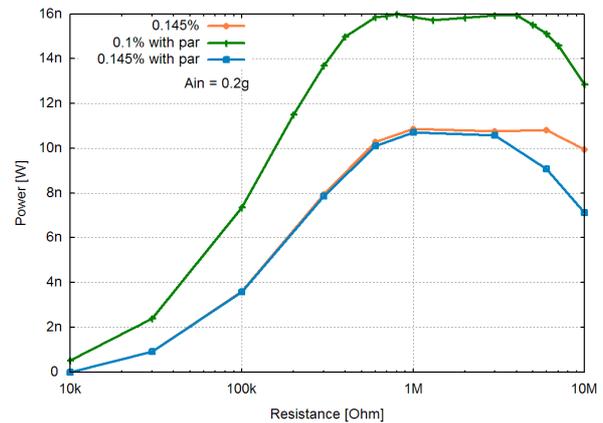

*Figure 3: FEM simulation results for the output power versus the load resistance value.*

The parasitic capacity decreases the coupling coefficient of the structure but not the output power. In fact the coupling coefficient of the structure is higher than the critical coupling after which the output power ceases to increase. It can be stated from the existence of two peaks in the frequency domain. We found that the value of frequency for which the power is maximal changes between 1577.5Hz and 1581.5Hz respectively for low load resistances and high load resistances.

## 2.3. POWER MANAGEMENT CIRCUIT

We have analysed two solutions for the rectification of the generated low amplitude signals. The first one consists in using discrete Schottky diodes and the other including a custom ASIC circuit.

## 2.4. Schottky diode rectifier

The rectifier circuit used is shown in the Figure 4.

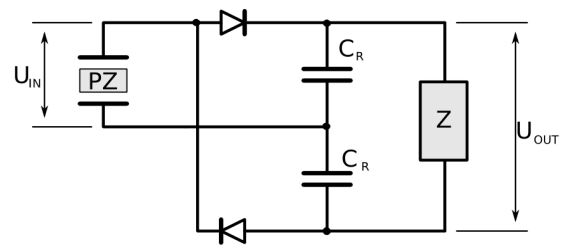

*Figure 4: Schematic view of the voltage doubling rectifier used with $C_R = 6.8nF$ and load impedance Z.*

It is composed of two HP 5082-2835 Schottky diodes and two 6.8nF capacitors. The circuit is capable of doubling the input voltage amplitude degraded by the threshold voltage of the diodes. Even if the input voltage is sufficiently high to surpass the threshold voltage it is very important to use diodes with this value the lowest possible in order to limit the power lost on these elements.

## 2.5. Custom ASIC circuit

In order to increase the voltages delivered by the circuit we propose to use a voltage multiplier circuit based on Villard charge pump structure (Figure 5).

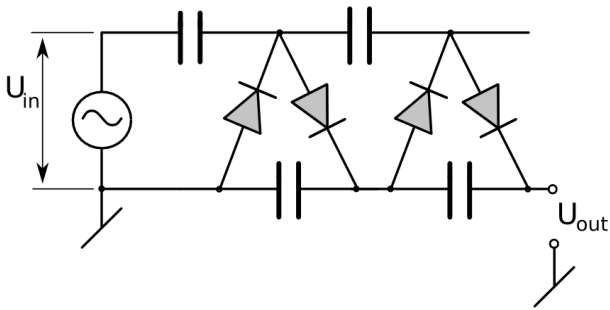

*Figure 5: Conventional structure of 4 stage voltage multiplier (Villard structure).*

Furthermore we propose to implement this circuit using very low threshold voltage diodes based on DTMOS transistors. The resulting circuit is realised as an ASIC in the 0.12μ HCMOS9 technology from ST microelectronics and occupies 1mm$^2$. For the test we used the circuit in a DIL8 package, but in the future we plan to connect directly the two chips. It contains six levels with low threshold voltage diodes and MIM capacitors. In order to optimise the efficiency of the system, the integrated capacitors value is 40pF, equal to the capacity of the MEMS generator. It rectifies and boosts the input voltage and can use input voltages of very low amplitudes.

## 3. EXPERIMENTAL RESULTS

We have characterised the circuit in laboratory environment. A DataPhysics Signal Force V20 shaker was used to generate input vibrations and a custom LabVIEW application was controlling the data acquisition.

## 3.1. Schottky diodes

The Figure 6 shows a comparison between the output voltage amplitudes and powers for three cases. The first one presents the open circuit voltage generated by the MEMS transducer. The second one uses a matched resistive load of 430kΩ connected directly to the piezoelectric generator, thus amplitude of alternative signal is presented and the RMS power dissipated on the load resistor. Finally the third uses the rectification circuit with Schottky diodes and a 10MΩ resistor as the Z load, so the presented signal is DC. The voltage amplitude in the second case is much higher than the others because of the use of voltage doubler and high load resistance value. On the other hand the output power is higher in the first case because there is no loss on the diodes. The power levels are very close however, 580nW for the rectified signal and 700nW for the alternative one.

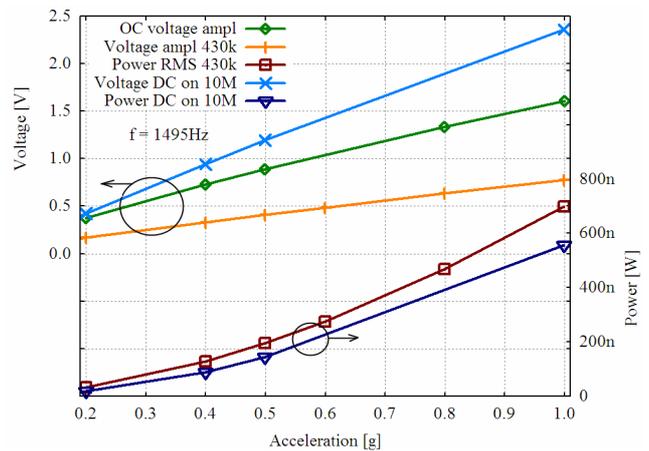

*Figure 6: Experimental output voltages and powers for three cases: open circuit, a matched 430kΩ resistor connected directly to the generator and a 10MΩ load connected through the rectification circuit.*

The proposed circuit will be used to charge a storage element which will power a wireless sensor node. The Figure 7 shows voltage curves versus time for charging a 1μF capacitor through the proposed rectification circuit. The generator was driven at resonance (1495Hz) with different excitation levels.

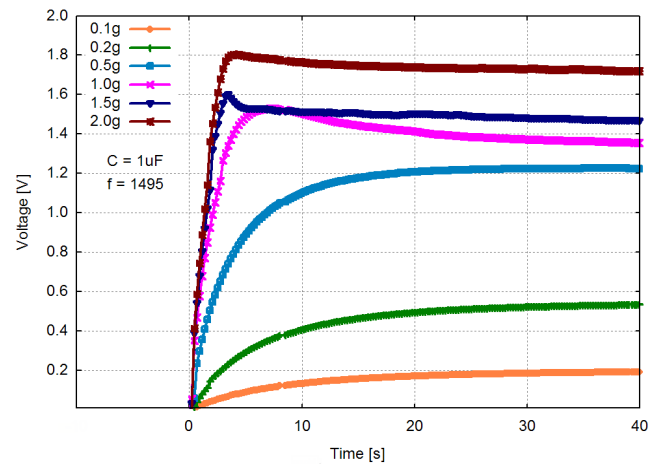

*Figure 7: Voltage on load capacitor of 1uF versus time for different input acceleration levels.*

The Figure 8 shows the maximum peak to peak values of the input voltage generated by the MEMS generator, the maximum voltage obtained on the load capacitor and the maximal instantaneous power of charging versus acceleration amplitude.

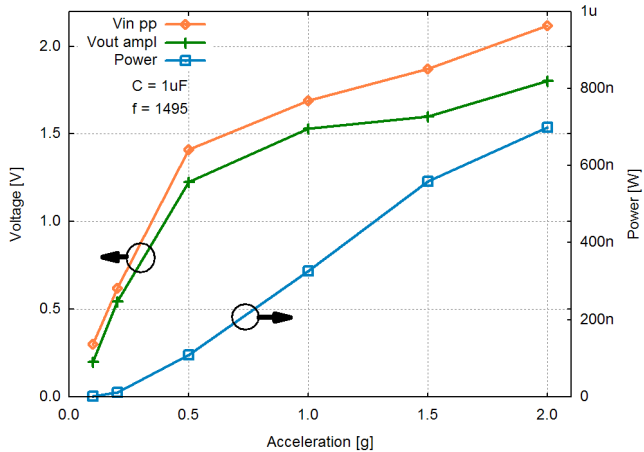

*Figure 8: Input and output voltages for charging a 1uF load capacitor as well as the maximum instantaneous power versus input acceleration value.*

The output voltage does not increase linearly with the input acceleration and reaches almost 2V for 2g excitation. The maximum instantaneous power of charging of the capacitor was calculated from temporal variation of the energy stored on this capacitor. It reaches 300nW for 1g excitation, which is at least half less than in the case of resistive load. It is caused by the nature of capacity to capacity energy transfer which implies that at least half of the energy must be lost.

### 3.2. ASIC circuit

The Figure 9 shows the relation between the output voltage generated by the VM circuit versus the input voltage provided by the MEMS generator.

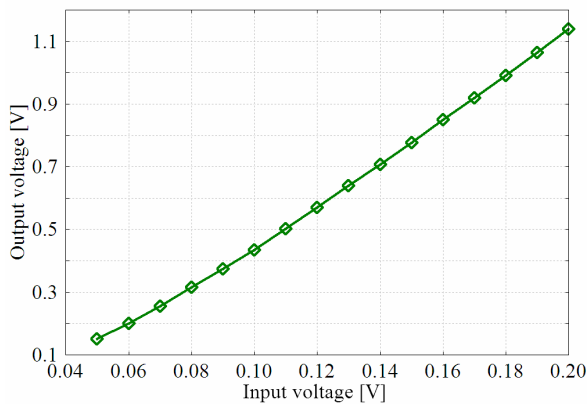

*Figure 9: Output voltage produced by the VM circuit versus the input voltage provided by the micro generator.*

A multiplication factor of 5.5 and of about 3 for low input voltages can be observed.

Thanks to the high multiplication ratio, the system is capable of generating high output voltages even at very low input voltages. The use of very low threshold voltage diodes permits to rectify voltages as low as 50mV in amplitude, which was impossible with the previously analysed circuit with Schottky diodes. However, the power efficiency of this circuit needs to be further analysed.

### 4. CONCLUSIONS AND FUTURE WORK

We present in this paper an analysis of mechanical vibration energy scavenging system. It uses a MEMS piezoelectric transducer and power management circuit to charge an energy storage unit. Two approaches were explored: one consisting of discrete Schottky diodes and the other using a custom ASIC. We have presented experimental results showing the capacity of this system to charge storage elements even from very low input vibration signals. It is an innovative solution for powering autonomous systems, for example Wireless Sensor Nodes.

The future work consists in using other piezoelectric materials for the MEMS transducer and optimising the composition of the power management circuit. Furthermore we plan to implement the system as a System On Package (SoP) in order to reduce its size and cost of fabrication.